# TITLE: AUDITORY CONVERSATIONAL BAI: A FEASIBILITY STUDY

Michal Robert Žák[1,2], Moritz Grosse-Wentrup[1,3,4]

We introduce a novel auditory brain-computer interface (BCI) paradigm, Auditory Intention Decoding (AID), designed to enhance communication capabilities within the brain-AI interface (BAI) system EEGChat [1]. AID enables users to select among multiple auditory options (intentions) by analyzing their brain responses, offering a pathway to construct a communication system that requires neither muscle movement nor syntactic formation. To evaluate the feasibility of this paradigm, we conducted a proof-of-concept study. The results demonstrated statistically significant decoding performance, validating the approach's potential. Despite these promising findings, further optimization is required to enhance system performance and realize the paradigm's practical application.

**INTRODUCTION:**
Brain-computer interfaces (BCIs) enable non-muscular interaction with the world, offering benefits for various populations [2]. Passive BCIs [3] and reactive BCIs, in particular, hold promise for creating "BCI companions" that assist users in daily life without requiring explicit control. Recently, a new extension of BCI technology was introduced: the brain-AI interface (BAI) [1]. This innovative approach integrates an AI agent into the BCI pipeline, enabling a new class of applications.

An example of a BAI is EEGChat, also introduced by [1], a conversational system designed to leverage large language models (LLMs) to restore lost linguistic functions to people affected by conditions such as aphasia. In a typical interaction with EEGChat, an interviewer might ask, "Hello, this is Pizzeria Romano, how can I help you?" EEGChat then generates and presents a list of keywords—such as "Menu," "Hours," "Order," and "Reserve"—to the subject. The subject focuses on one keyword, such as "Reserve," and EEGChat recognizes their selection. Finally, EEGChat expands the chosen keyword into a full-sentence response, such as "Hello, I would like to make a reservation," and reads it out loud.

EEGChat operates using the code visually evoked potential (cVEP) paradigm, wherein answer options utilize differently flashing patterns that can be detected in the subject's brain responses. While effective, visual stimuli have inherent limitations, such as requiring sustained visual attention, which hinders natural interaction and restricts usability in most environments.

To address these challenges, we propose a novel auditory stimulation-based reactive BCI paradigm termed Auditory Intention Decoding (AID). This paradigm is designed to enhance EEGChat's functionality by allowing users to implicitly communicate intentions through auditory stimulation, without requiring explicit physical or linguistic actions. For instance, in an interaction with EEGChat-AID, the interviewer might ask, "Hello, this is Pizzeria Romano, how can I help you?" EEGChat-AID would then extend this questions with possible answer options, such as "[…] Can I take your order, reserve a table, or show you the menu?" Based on the subject's brain responses to each option, EEGChat-AID would identify their intention—for example, to reserve a table—and expand it into a full-sentence answer, such as "Hello, I would like to make a reservation," which is then read aloud.

By eliminating the reliance on visual attention, AID provides an appealing solution for real-time,

[1]Research Group Neuroinformatics, Faculty of Computer Science, University of Vienna, Vienna, Austria
[2]Doctoral School Computer Science, Faculty of Computer Science, University of Vienna, Vienna, Austria
[3]Vienna Cognitive Science Hub, University of Vienna, Vienna, Austria
[4]Research Network Data Science, University of Vienna, Vienna, Austria

accessible communication, particularly for individuals with severe speech or physical disabilities. This auditory paradigm paves the way for a more intuitive, unobtrusive alternative to traditional visual-based BCIs, unlocking new possibilities for future BCI-driven interaction.

**RELATED WORK:**

AID represents a novel paradigm that, in its current conceptualization, appears to be absent from existing literature. While auditory brain-computer interfaces (BCIs) have been studied, a lot of the research has focused on source selection—determining which of several auditory streams a user is attending to [4, 5, 6]. Reactive BCIs have also been explored, particularly those leveraging the N400 event-related potential (ERP) [7, 8], which has demonstrated measurable effects in auditory processing experiments.

A closely related study to AID by [9] investigated the feasibility of using an N400 ERP-based BCI to decode concepts from a user's thoughts. Participants were presented with words from different semantic categories, and their neural responses were analyzed to infer the category of focus. Although the study confirmed the presence of an N400 effect, the strength of the signal and its decoding reliability were limited. Given that this work was conducted seven years ago and no significant follow-up studies appear to have expanded on this approach, we recognize the need to revisit such paradigms. By applying more advanced methodologies and tools, AID aims to address these limitations.

**EXPERIMENT AND RESULTS:**

To assess the feasibility of the AID paradigm, we conducted a proof-of-concept experiment focusing on the decoding component. At this stage, before integrating a full language model, it was critical to establish a functional framework capable of decoding user intentions. To simplify the study while preserving the paradigm's integrity, three three-digit numbers were used as auditory stimuli instead of words.

Each trial followed a structured protocol: participants were primed with a specific three-digit number, after which they sequentially heard three numbers, one of which matched the primed number. The experiment included 20 rounds, each consisting of 8 trials, yielding a total of 160 usable trials per participant. Short breaks between rounds were incorporated to mitigate fatigue and preserve data quality.

To decode complex structures in EEG signals and assess system performance, we utilized EEGNet [10]. EEG samples were windowed between -0.2 and 1.85 seconds relative to stimulus onset and labeled as target (primed) or non-target (non-primed). Due to the experimental design, two-thirds of the data represented non-target instances, which influenced the statistical significance threshold. EEGNet was configured with the Adam optimizer, a learning rate of 0.001, and a batch size of 350 and run with a ten fold cross-validation (CV).

Preliminary results demonstrated that seven out of ten participants achieved statistically significant classification accuracy, as assessed via a permutation analysis (n = 10,000). Mean accuracies across CV folds ranged between 0.6 and 0.7, with standard deviations of approximately 0.04. While these results are preliminary, they clearly indicate the presence of an elicited effect in the EEG data. Future improvements, such as optimizing the temporal window and fine-tuning EEGNet parameters, hold the potential to further enhance classification performance.

**CONCLUSION:**

The potential of an AID-like paradigm highlights the need for continued research to build upon the foundation established in this study. While our initial decoding attempt demonstrated some success, the achieved accuracies are not yet sufficient for practical application. Current results indicate that multiple repetitions of stimuli may be required to achieve the precision necessary for reliable use—an approach that conflicts with the paradigm's envisioned seamless functionality.

By refining the decoding framework and addressing its limitations, future research could unlock the full potential of this paradigm, leading to the development of a novel communication tool. Such a tool could significantly enhance the quality of life for individuals with severe language impairments, providing them with an effective means of expression and interaction.